\documentclass[aps,12pt]{revtex4}

\usepackage{graphicx}        

\usepackage{bbm,mathrsfs,indentfirst,nicefrac,units}
\usepackage{times}
\usepackage{float}
\usepackage{amsmath}
\usepackage{amsfonts}
\usepackage{amssymb}
\usepackage{epsfig}
\usepackage{graphicx}
\usepackage{dcolumn}
\usepackage{bm}

\usepackage{verbatim} 
\usepackage{subfig}
\newcommand{\half}{\frac{1}{2}\;}
\newcommand{\ns}{n_s}
\newcommand{\nsp}{n_{s'}}
\newcommand{\nssp}{n_{s-s'}}

\newcommand{\fr}{\mathcal{FR}}
\newcommand{\frr}{\mathcal{F}}
\newcommand{\co}{\mathcal{C}}
\newcommand{\fp}{\widetilde{\mathcal{F}}}
\newcommand{\pfr}{f}

\begin{document}

\title{Relating the microscopic rules in coalescence-fragmentation models\\
to the cluster-size distribution}
\author{B\l a\.{z}ej Ruszczycki$^1$, Ben Burnett$^2$, Zhenyuan Zhao$^1$, and Neil F. Johnson$^1$}
\affiliation{$^1$Physics Department, University of Miami, Coral Gables,
FL 33146, USA \\
$^2$Rudolf Peierls Centre for Theoretical Physics, Oxford University, Oxford, U.K.}

\date{\today}

\begin{abstract}
Coalescence-fragmentation problems are now of great interest across the physical, biological, and social sciences. They are typically studied from the perspective of rate equations, at the heart of which are the rules used for coalescence and fragmentation. 
Here we discuss how changes in these microscopic rules affect the macroscopic cluster-size distribution which emerges from the solution to the rate equation. Our analysis elucidates the crucial role that the fragmentation rule can play in such dynamical grouping models.
We focus our discussion on two well-known models whose fragmentation rules lie at opposite extremes. In particular, we provide a range of generalizations and new analytic results for the well-known model of social group formation developed by Egu\'iluz and Zimmermann [V. M. Egu\'iluz and M. G. Zimmermann, Phys.
Rev. Lett. \textbf{85}, 5659 (2000)]. 
We develop analytic perturbation treatments of this original model, and extend the analytic analysis to the treatment of growing and declining populations.  
\end{abstract}


\maketitle

\section{Introduction\label{sec:Introduction}}
The challenge to understand the dynamics of Complex Systems is attracting increasing attention, particularly in the socio-economic and biological domains \cite{fjmo,ez,gl,neil,neil1,clauset1,neil2,marker2,marker3,mhj1,mhj2,marker4,marker5,marker6,marker7,marker8,marker9,marker10,marker13,
marker14}.  For example, the recent turmoil in the financial markets has created significant public speculation as to the root cause of the observed fluctuations. At their heart, all Complex Systems share the common
property of featuring many interacting objects from which the
observed macroscopic features emerge. Exactly how this happens
cannot yet be specified in a generic way -- however, an important milestone in this endeavor is to develop a quantitative understanding of any internal clustering dynamics within the population.
Coalescence-fragmentation processes have been studied widely in
conventional chemistry and physics \cite{smoluchowski1,smoluchowski2,leyvraz0,leyvraz1,drake,aldous,laurencot0,laurencot1,marker15,marker16,marker17,marker18,
marker19,marker20,marker21,marker22,marker23,marker24,marker25,marker26,marker27,bc}
 -- however, collective behavior
in social systems is not limited by nearest neighbor interactions,
nor are the details of social coalescence or fragmentation
processes necessarily the same as in physical and biological
systems. The challenge for a theorist is then twofold: (1) to provide a model which accounts correctly for the observed real-world behavior --- e.g., in the case that power-laws are observed empirically, the model should be able to reproduce the power-law dependence itself, the value of the corresponding power-law exponent, and possibly also the form of the truncation; (2) the rules invoked in the model need to make sense in the context of the real-world system being discussed.

In this paper, we discuss coalescence and fragmentation problems with a focus on social
systems. In particular, we consider interactions which are essentially independent of spatial separation in order to mimic the effect of modern communications etc. Much of our discussion is focused around fragmentation processes in which an entire cluster breaks up into its individual pieces -- thereby mimicking a social group disbanding --
as opposed to the more typical case studied in physical and biological systems of binary splitting. We limit our discussion to the steady-state behavior corresponding to a constant population, or a steadily growing/declining population. In Sec.~\ref{gf}, we lay out a general formulation of such coalescence-fragmentation problems. In order
to understand the quantitative effects of a particular choice of fragmentation rule, Sec.~\ref{rff} then compares two well known
coalescence-fragmentation models, with fragmentation rules which lie at opposite extremes of the spectrum.
 One of these is the well-known physics-inspired model of social group formation introduced by Egu\'iluz and Zimmermann \cite{ez} while the other is a standard model in mathematical ecology due to Gueron and Levin \cite{gl}. The explicit comparison between the two models allows us to elucidate the subtle differences in their microscopic rules that make their macroscopic distributions differ, and leads us to a better generic understanding of the crucial role that the fragmentation rule can play. We then proceed to focus on the physics-based model of Egu\'iluz and Zimmermann, generalizing it in several ways and providing new analytic results (Sec.~\ref{sec:Generalization}). We analyze a perturbed version of the Egu\'iluz-Zimmerman model where spontaneous cluster formation is present (Sec.~\ref{prtb}), as well as generalized versions in which there is a steadily growing 
(Sec.~\ref{sec:increasing}) or declining population (Sec.~\ref{sec:decreasing}). Further realistic modifications of the Egu\'iluz-Zimmerman model are discussed in Sec.~\ref{modfs}.

There is of course a huge volume of work in the mathematics, physics and
chemistry literature on the topic of clustering within a many-body
population of interacting particles \cite{wattis}. The Smoluchowski
coalescence equation is arguably the most famous and well-studied
example \cite{smoluchowski1,smoluchowski2,leyvraz0}. Reference \cite{wattis} provides an excellent recent review of coalescence-fragmentation models in physical and chemical systems from a  mathematician's perspective -- however we note that the socially-inspired models that we focus upon in this article are not discussed. Many previous studies have tended to focus on generic  mathematical issues such as
existence, uniqueness, mass conservation, gelation and finite size
effects (see Refs.\cite{drake,aldous,leyvraz1,laurencot0,laurencot1} and references
therein). When it comes to Complex Systems -- and in particular,
social systems -- the more pressing goal is to understand the
emergent features of the population. In contrast to physical and
chemical systems in which collision energetics play a crucial role
in guiding the specification of microscopic coalescence and
fragmentation rules, the precise microscopic rules in social
systems are unknown -- however, the overall macroscopic emergent
phenomena such as cluster size distribution can be measured
relatively easily. In financial markets, the
collective dynamics of the population of traders is registered
directly by means of the price. Indeed, as many prior works have
shown, such collective behavior in social systems tends to produce
near scale-free (i.e. power-law) networks and/or cluster sizes in a
variety of real-world situations. For example, the distribution of transaction sizes follows a power-law with
slope near 2.5 for each of the three major stock exchanges in New York, Paris and London \cite{gabaix}. In addition, it has been shown that the distribution of the severity of violent events inflicted in conflict by insurgent groups, and by terrorist groups, follow a power-law near 2.5 \cite{neil1,clauset1,neil2,clauset2}. 
The model of Egu\'iluz and Zimmermann \cite{ez}, which is the starting point of much of the paper's discussion, is therefore an attractive candidate model for such social systems. In addition to its intrinsic theoretical interest because of its non-binary fragmentation rule, which mimics the disbanding of social groups, it also happens to produce a robust power-law distribution of cluster sizes with slope 2.5 \cite{ez}.

\begin{figure}
\includegraphics[width=0.5\textwidth]{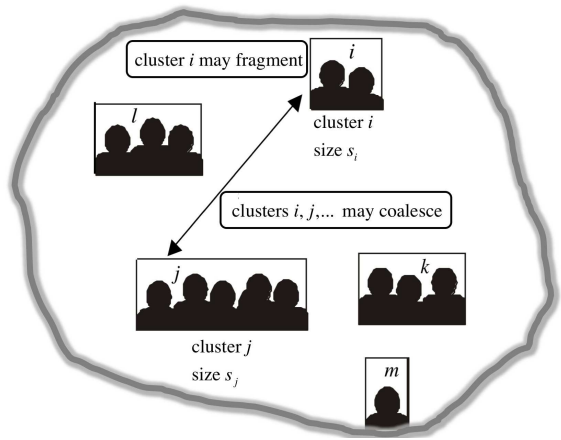}
\caption{Schematic diagram indicating the presence of  coalescence
and fragmentation processes, for a population of $N=15$ objects
dynamically partitioned into clusters. The size of cluster $i$ is
$s_i=2$, while the size of cluster $j$ is $s_j=6$ etc. The
fragmentation process exhibits the richest range of
possibilities, given the combinatorial number of ways in which a cluster can in principle be divided. There are many possible realizations of the objects
themselves, e.g.  humans, animals, macromolecules, though for
simplicity we show them as humans.}\label{fig:groups}
\end{figure}

\subsection{Modeling Social Systems\label{sec:Introduction}}
Many social systems 
seem to comprise a large number of dynamically evolving clusters. Over time, and in an apparently self-organized way, clusters either coalesce with each other to form even larger clusters, or fragment to form a collection of smaller ones. In addition to everyday social situations, these characteristics seem consistent with common sense notions of the dynamical connectivity within a community of financial traders \cite{ez}, or even a loosely connected insurgent population or terrorist/criminal network \cite{neil1,neil2}. Figure~\ref{fig:groups} illustrates the generic situation of interest in many  recent works on  coalescence-fragmentation models \cite{ez,gl,marker2,marker3,mhj1,mhj2,marker4,marker5,marker6,marker7,marker8,marker9,marker10,marker13,marker14,neil}. As a result of coalescence and fragmentation processes over time, the population of $N$ objects undergoes dynamical partitioning into clusters $i, j, k, \dots$ of size $s_i, s_j, s_k, \dots$, where both the number of clusters and their membership are typically time-dependent.  We have denoted the $N$ objects in human form, but of course they could be animals, macromolecules or other indivisible entities. Earlier studies tended to focus on situations in which the interactions between clusters might be expected to decay with physical separation -- as in a simple solution of molecules interacting through Van der Waals interactions for example. However in modern-day social applications, where long-distance communication is as commonplace as communication with neighbors, it makes more sense to have interactions over all lengthscales, with the interaction probability effectively independent of physical separation. These are the type of interactions that we discuss here.

Of the two processes in Fig. 1, i.e. coalescence and fragmentation, the coalescence process is likely to be the simpler and more generic. Suppose we have a particular partition of a population of $N$ objects into clusters as in Fig. 1, and that a cluster $i$ of size $s_i=2$ is to coalesce. It is unlikely to undergo three-body collisions and/or interactions, and hence its most likely coalescence event is to join with a single other cluster $j$. Given that the size of a cluster measures the number of objects in it, it is therefore reasonable to imagine that the coalescence probability should increase as the size of the clusters themselves increase. In a more human setting, the more objects that a cluster contains, the more likely it is that something will happen to one of its members in order to induce such an event, and hence the probability will increase with the size of the cluster. We therefore adopt size-dependent coalescence probabilities in this work. We note that although we are using the term `cluster' throughout this paper for convenience, it can also be taken to mean a `community' in the language of network science\cite{book} since it denotes a subset of the population who have very strong links between them, while the links between clusters are negligibly weak. We note also that the term `cluster' need not necessarily mean physical connection -- instead it could represent a group of objects whose actions happen to be coordinated in some way. Hence the coalescing of two clusters, however distant in real space, can mean an instantaneous alignment of their coordinated activities, as one might expect in a financial market\cite{neil}, organized crime or insurgent warfare\cite{neil1,neil2}. In such a situation, a common fragmentation event would then likely be a sudden disruption of this coordination -- hence it is this type of fragmentation rule that forms the focus of our work. Although we do not explore the details of real-world applications such as financial markets or insurgent warfare here, it is useful to keep them in mind when we discuss the consequences of the different fragmentation rules later in the paper.

As mentioned earlier, the distinct feature of many real-world systems is the
existence of scale-free behavior in the time-averaged cluster size distribution\cite{ez,neil,neil1,neil2,newman,econo,gabaix}, such that in the first instance these systems can be 
characterized by the exponent of their power law and by the range of its scale-free behavior. One may therefore ask: Which ingredients of the coalescence-fragmentation models, or combinations of ingredients, turn out to control the various observable aspects? It is this general question that motivates the present work.

\subsection{General Formulation}\label{gf}
Once the probabilities specifying the coalescence and fragmentation are given, the cluster size distribution may be computed either
by a direct simulation of the model or in a mean-field theory approximation by solving an appropriate set of rate equations, often numerically. 
The rate equations are typically non-linear. The non-trivial question 
of existence and uniqueness of the time-independent solution therefore arises, and is addressed in seminal works such as Refs. ~\cite{bc,cc1,cc2}.
For the social/economical models of current interest, the uniqueness and existence can be shown at the level of the rate equations, and verified by direct simulations.
We consider mostly `steady-state' models, in which there is some form of robust long-time behavior. 

\begin{figure}[ht]
\centering
\includegraphics[width=0.8\textwidth]{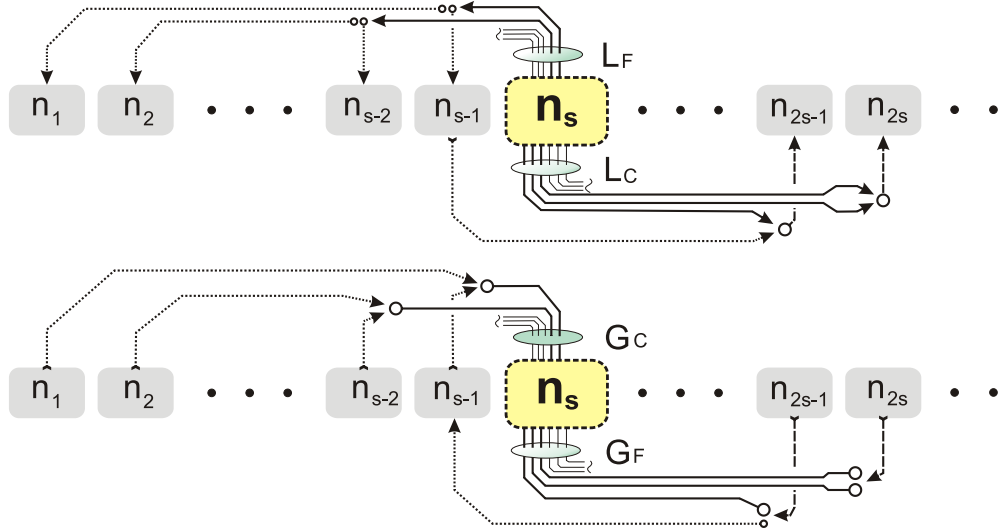}
\caption{The various processes of cluster coalescence and fragmentation which give rise to $L_F$, $L_C$, $G_F$, $G_C$ for 
any particular value of $s$. The top figure represents the appearance of new clusters of size $s$, the bottom one represents their loss.
In the interests of simplicity, the fragmentation into two clusters has been depicted and only a few processes are shown.}\label{fig:1}
\end{figure}

The number of clusters of size $s$ at time $t$ is $\ns(t)$, and $N$ is the total number of members 
(i.e. the population size). We will drop the explicit time-dependence for simplicity, since it will be clear from the context whether we are discussing  $\ns(t)$ or its steady-state time-averaged value. In order to characterize a general system, we need to prescribe the following two functions, each of dimension ${\rm [time]}^{-1}$:
\begin{itemize}
\item The {\it coalescence function} $\co(s,s')$ which is the rate describing the process by which
two clusters of sizes $s$ and $s'$ merge.
We only consider coalescence which depends on the details of a pair of clusters, and hence exclude the possibility that 3 (or more) clusters are involved in the merging process.
\item The {\it fragmentation function} $\fr(s;m_1,m_2,\ldots,m_n-1)$ which is the rate describing the process by which a cluster of size $s$ fragments
into a configuration which contains $m_1$ clusters of size 1, $m_2$ clusters of size 2, etc.
\end{itemize}
The functional form of the above two functions is taken to be time-independent.
If we consider general fragmentation processes, we see that a large number of parameters are necessary to characterize the
fragmentation.  
However in order to write down the rate equations and hence calculate the cluster size distribution, we do not need complete knowledge of the fragmentation function (i.e. we do not need knowledge about all possible partitions).
It is sufficient to know the \textit{reduced fragmentation function} $\frr(s,s',m)$, defined as 
the rate at which a cluster of size $s$ fragments
into a configuration which contains $m$ clusters of size $s$ plus any other clusters with sizes different to $s'$. In
addition to $\frr(s,s',m)$ we need to know the rate that 
the fragmentation of any given cluster of size $s$ occurs, which we denote as $\pfr(s)$. In principle 
we can calculate it by summing the complete fragmentation function over all partitions of the fragmentation products. 
By prescribing the deduced fragmentation function $\frr(s,s',m)$ we do not characterize 
uniquely the fragmentation of the system and in general we may not be able to calculate $\pfr(s)$ -- yet it is possible in specific cases to do so
once the assumption regarding the fragmentation products has been stated.
Looking at the average number of clusters of size $s$ that in unit time undergo the various processes (see Fig. \ref{fig:1}), 
we may introduce the following notation:
\begin{itemize}
\item
$L_F(s)$: {\it loss due to fragmentation}, the number of clusters of size $s$ that fragment

\item
$L_C(s)$: {\it loss due to coalescence}, the number of clusters of size $s$ that join with other clusters

\item
$G_C(s)$: {\it gain from coalescence}, the number of clusters of size $s$ created from the merging of clusters of size smaller than $s$

\item
$G_F(s)$: {\it gain from fragmentation}, the number of clusters of size $s$ created from fragmenting clusters of size larger than $s$
\end{itemize}
Symbolically the rate equations for any $s$ are written as
\begin{equation}
\frac{\partial\,n_s}{\partial\,t}=-L_F(s)-L_C(s)+G_C(s)+G_F(s)\label{eqcon}
\end{equation}
which explicitly reads as
\begin{equation}
\frac{\partial\,n_s}{\partial\,t}=-\pfr(s)\,\ns-\ns \sum_{s'=1}^{N} \nsp \co(s,s')+\half \sum_{s'=1}^{s-1} \nsp \nssp \co(s',s-s')+
\sum_{s+1}^N\nsp \, \sum_{m=1}^{[N/s']}m\,\frr(s',s,m)\,.\label{meq}
\end{equation} 
The last term represents the gain in the number of clusters of size $s$ coming from fragmentation of other clusters of size $s'>s$,
in such a way that among the fragmentation products we have $m$ clusters of size $s$. We are summing over all possible values 
of $m$ and $s'$. Note that we sum over $s'$ which is here the first (not the second) argument of $\frr$.
An explicit form for $f(s)$ is discussed above. It is convenient to formally define  
\begin{equation}
\fp(s,s')=\sum_{m=1}^{[N/s']}m\,\frr(s,s',m)\,.\label{rdfrag}
\end{equation}
We write therefore the last term of Eq. \eqref{meq} as
\begin{equation*}
\sum_{s'=s+1}^N\nsp \fp(s',s)\,.
\end{equation*}

\section{Role of the fragmentation function}\label{rff}
A logical first step in the quest to understand classes of models which differ in their cluster fragmentation process,  is to look at extreme cases. One such case is the Egu\'iluz-Zimmermann (E-Z) model~\cite{ez}. In the E-Z model,  
fragmentation of a cluster of size $s$ always produces $s$ clusters of size $1$, i.e. the cluster breaks up into individual objects. At the other extreme, is the famous Gueron and Levin (G-L) model~\cite{gl} in which fragmentation of a cluster yields two smaller pieces, i.e. the original cluster splits into two clusters. 
The original G-L model is formulated
in terms of continuous distributions -- however, since our aim is to analyze the effects of these rules on the same footing, we will focus on the discrete version of the G-L model, returning to the continuous formulation later on.
The G-L model is in fact identical to Smoluchowski's coagulation-fragmentation model with binary fragmentation.

The common feature of the models that we discuss, is the presence of a separable coalescence function: 
\begin{equation}
\co(s,s')=\alpha\,a(s)\,a(s')\,\ .
\end{equation}
In principle, the multiplicative constant may be absorbed into $a(s)$, however we prefer to keep it explicitly and 
adopt a dimensionless $a(s)$. 
This class of model is further specified by introducing a coalescence mechanism on the microscopic scale, namely that two clusters merge 
when any member from one cluster connects to any member from the other cluster. In a macroscopic description, this is equivalent to assuming that $a(s)=s$. We note that Gueron and Levin\cite{gl}, having the solution of the rate equations for $a(s)=s$, considered explicitly the other cases $a(s)=1$ and $a(s)=1/s$  by means of the substitution $n_s \rightarrow a(s)\,n_s$ -- however, this substitution affects the form of the fragmentation function $\frr(s,s',m)$.

\subsection{Fragmentation function}
Assuming that the cluster may only split into two pieces still does not uniquely specify the fragmentation, since we still need information
about the probability distribution for the sizes of the fragments. In the G-L model, it is stated that
the conditional distribution for fragments is uniform\cite{gl}, i.e. the fragmentation of a cluster occurs with a probability which is 
independent of the way in which the cluster breaks. The reduced fragmentation function for $s>1$ is therefore 
\begin{equation}
\frr_{GL}(s,s',m)=\beta\, b(s) \left[2\,\delta_{m,1}(1-\delta_{2s',s})+\delta_{m,2}\delta_{2s',s}\right]\, \label{GL_frag}
\end{equation}
where we have accounted for the fact that if $2s'=s$, the cluster breaks into two fragments of equal size. 
Using Eq. \eqref{rdfrag} one obtains immediately 
\begin{equation}
\fp_{GL}(s,s')=2\beta\,b(s)\, \ \ .
\end{equation}
The fragmentation probability is calculated as follows: 
\begin{equation}
f_{GL}(s)=\half\sum^{s-1}_{s'=1} \,\frr_{GL}(s,s',m=1)+\sum^{s-1}_{s'=1}\,\frr_{GL}(s,s',m=2)=\beta \,(s-1) \,b(s)\,,
\end{equation}
where the factor $\nicefrac{1}{2}$ in the first term appears in order to avoid double-counting, and the second term represents splitting into two equal parts.  
In the E-Z fragmentation scheme, the cluster of size $s$ can only break up into individual objects and there is only one mode of 
fragmentation, hence 
\begin{equation}
\frr_{EZ}(s,s',m)=\beta\, b(s)(1-\delta_{s1}) \delta_{s',1}\delta_{m,s}\,.
\end{equation}
Using Eq. \eqref{rdfrag} we have
\begin{equation}
\fp_{EZ}(s,s')=\beta s\, b(s)\,(1-\delta_{s1})\delta_{s',1} \,.
\end{equation}
The fragmentation probability is 
\begin{equation}
f_{EZ}(s)=\sum_{s'=1}^{s-1}\,\frr_{EZ}(s,s',m=s)=\beta\,(1-\delta_{s1}) b(s).
\end{equation}
There is no double-counting problem here.
A peculiar feature of the E-Z model is that the corresponding set of rate equations is semi-recursive, i.e. any $k$-th equation depends only
on values of $n_{s'}$ for $s' \leq k$ and on a global constant depending on all $n_s$.  This is a feature by which it is easy to
show the existence and uniqueness of the solution and also to solve the system
numerically. 

It is the common feature of both G-L and E-Z type models to assume that $a(s)=b(s)$.
Mathematically, this acts to restrict the space of all possible solutions, otherwise the diversity of general solutions would be overwhelming. In physical terms, the justification for this assumption is that there is interest in the specific case $a(s)=b(s)=s$, since this describes the case of fragmentation of the cluster being triggered by a single member -- hence the proportionality to  $s$. Similarly the likelihood that two groups would become coordinated and hence act as a single unit (i.e. they coalesce) would be proportional to the size of each of the groups, if the underlying mechanism involved one member from each initiating the process by forming a link followed by all the other members.

With the assumptions made so far, it turns out that each system is described by three constants: $\alpha$, $\beta$ and the total population 
size $N$. For the time-independent system we need just two constants, and since $\alpha$ and $\beta$ are of dimension ${\rm [time]}^{-1}$ then only their ratio $\alpha/\beta$ should appear.
The steady-state rate equations are as follows.
\begin{align}
\text{G-L system:}\notag\\
-\beta(s^2-s)\,&\ns-\alpha\,s\,\ns \sum_{s'=1}^N\,s' \nsp +\frac{\alpha}{2}\,\sum_{s'=1}^{s-1} \,s'\,\nsp \,(s-s')\nssp +
\;2\,\beta\,\sum_{s'=s+1}^Ns' \, \nsp =0\,.\label{glmeq}\\
\text{E-Z system:}\notag\\
-\beta s\,(1-\delta_{s1})&\ns-\alpha\,s\,\ns \sum_{s'=1}^N\,s' \nsp +\frac{\alpha}{2}\,\sum_{s'=1}^{s-1} \,s'\,\nsp \,(s-s')\nssp +
\beta\delta_{s,1}\sum_{s'=s+1}^N s'^2 \, \nsp =0\,.\label{ezmeq}
\end{align} 
\noindent Egu\'iluz and Zimmermann\cite{ez} explicitly used the following constants: 
\begin{equation}
\alpha=\frac{2(1-\nu)}{N^2}\,,\qquad \beta=\frac{\nu}{N}\,.\label{ezconst}
\end{equation}
We see that both sets of equations \eqref{glmeq} and \eqref{ezmeq} simplify if 
we express them in terms of $k_s=s\,n_s$, i.e.
the number of agents contained in clusters of size $s$. Note that for general $a(s)$, we need to substitute $k_s=a(s)\,n_s$.

\subsection {Equilibrium in Gueron-Levin model: Continuous formulation}
Gueron and Levin's solution~\cite{gl} to the G-L model, was obtained for the system with continuous cluster density 
which we denote as $n(s)$. In terms of $k(s)=s\,n(s)$,
the integral rate equation corresponding 
to Eq. \eqref{glmeq}  with no limit on the maximum size of a cluster, is given by:
\begin{equation}
0=-\beta\,s\,k(s)-\alpha\,k(s)\int_0^\infty ds'\,k(s')+\alpha\half\int_0^s ds' k(s')\,k(s-s')+2\beta\int_s^\infty ds'\,k(s')\,.\label{contgl}
\end{equation}
Looking at this equation, we guess that the solution is obtained by substituting an ansatz which satisfies $k(s+s')\propto k(s)k(s')$.
The first form to try is $k(s)=A\,e^{-\mu\,s}$.
With this ansatz we obtain
\begin{equation}
0=-A\,\beta\,s\,e^{-\mu s}-A^2\alpha/\mu e^{-\mu s}+A^2\alpha/2\,s e^{-\mu s}+2A\beta/ \mu \,e^{-\mu s}\,.
\end{equation}
There are two types of terms, of the form $\sim e^{-\mu s}$ or $\sim s\,e^{-\mu s}$.
Eliminating the overall exponential factor we have
\begin{equation}
0=s\left(-A\beta+A^2\frac{\alpha}{2}\right)+\frac{2}{\mu}\left(A\beta-A^2\frac{\alpha}{2}\right)\,.
\end{equation}
Both terms in parentheses vanish if we choose 
\begin{equation}
A=2\frac{\beta}{\alpha}\,.
\end{equation}
The scale factor $\mu$ in the exponent is determined to be $\mu=\nicefrac{2\beta}{N\alpha}$ by normalization.
The solution to Eq. \eqref{contgl} is just an exponential function which was obtained by Gueron and Levin by means of a Laplace transform.

We notice here 
a remarkable curiosity: If we take the actual solution of Eq. \eqref{contgl}, then for any $s$ the following equalities hold exactly:
\begin{equation}
L_F(s)=G_C(s),\qquad L_C(s)=G_F(s)\,.
\end{equation} This is in effect the detailed balancing.
In other words, the following holds for the G-L model: {\em The average loss of clusters of size $s$ due to the cluster fragmentation, is equal to the average gain 
obtained from the coalescence of clusters of sizes smaller than $s$. 
Also the average loss of clusters of size s due the coalescence with other clusters
is equal to the average gain obtained from the fragmentation of clusters of sizes larger than s.}

In addition to its mathematical interest, this identity (which is not satisfied for the E-Z model as discussed below) shows up a fundamental feature of the G-L model, which  arises in turn from the microscopic rules which characterize it.  This symmetry is also revealed if we look at the behavior of the system with time flowing backwards.(In general, one does not obtain 
a stochastic system by time-reversing the recorded history of a second non-equilibrium stochastic system. Although 
this becomes an issue for discrete systems due to the presence of fluctuations, we may still discuss it from the perspective of the average quantities describing the equilibrium state). With the reversed time perspective,
the coalescence of clusters is observed as fragmentation and vice-versa, but the average cluster size distribution remains unaltered in the equilibrium state. As far as this  average quantity is concerned, the system is therefore invariant under an interchange of coalescence/fragmentation processes -- and in the specific case of G-L model, the time-reversed processes are exactly the same as the original ones.    

\subsection{Cluster size distribution: The exponential cutoff}
We now return to the discrete formulation.
For the discrete version of the G-L system, it may be
verified by direct computation that the steady-state value
\begin{equation}
n_s=2\frac{\beta}{\alpha}\,s^{-1}\exp\left(-\mu \,s\right)\,
\end{equation}
is also a solution of Eq. \eqref{glmeq}, once we make an approximation of extending the summation limits to infinity.
Here the normalization condition is 
$N=\sum_{s'=1}^\infty s'\,n(s')$, from which we calculate 
\begin{equation}
\mu=\ln \left(\frac{2\beta}{\alpha N} +1\right) \,.
\end{equation}
Thus we have 
\begin{equation}
n_s=2\frac{\beta}{\alpha}\,s^{-1}\,\left(\frac{2\beta}{\alpha N} +1\right)^{-s}\,.
\end{equation}
It is advantageous to consider $\beta / \alpha \propto N$, 
thus the exponent is independent of $N$ and $\ns$ is just proportional to $N$.
If we use here the same constants (Eq. \eqref{ezconst}) as the original E-Z model, the solution is
\begin{equation}
\text{G-L}:\;n_s=N\frac{\nu}{1-\nu}\,s^{-1}\,\left(1-\nu \right)^s\,.
\end{equation}
The solution to the E-Z model rate equations may be approximated as~\cite{marker3}
\begin{equation}
\text{E-Z}:\;n_s\sim N s^{-2.5}\left(\frac{4(1-\nu)}{(2-\nu)^2}\right)^s\,.\label{EZ1}
\end{equation}

In order to compare the cluster size distribution for both models, we will for convenience characterize both using the same
parameters $N$ and $\nu$. This means that they will have the same coalescence function, and their fragmentation functions will agree for the splitting of clusters of size $s=2$. The difference between the two models then lies in the fragmentation of larger clusters. This allows them to be compared on a similar footing, focusing just on the effect of their respective fragmentation functions.

\begin{figure}[ht]
\centering
\includegraphics[width=0.7\textwidth]{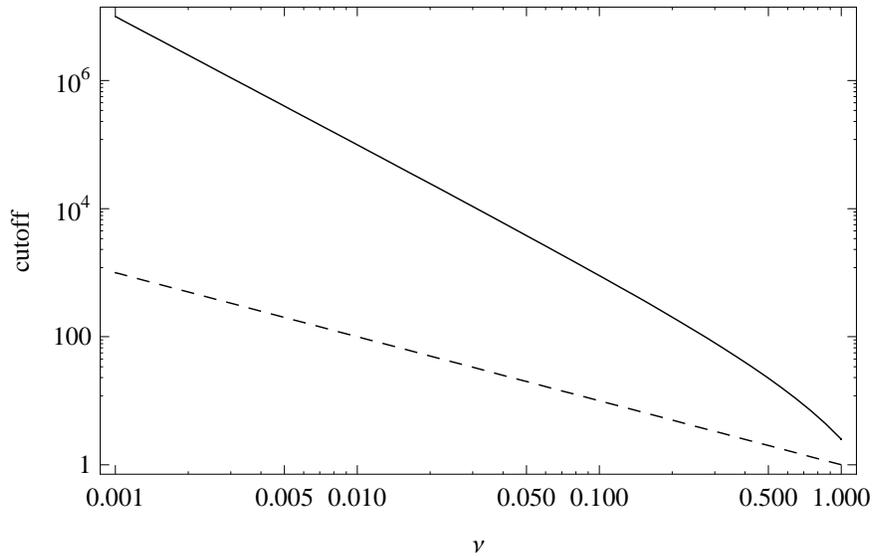}
\caption{Scale of exponential cutoff for the E-Z model (solid curve) and for the G-L model 
(dashed curve) described by the same parameter $\nu$. The range of cluster sizes for which one observes the power-law,
is several orders of magnitude larger for the E-Z model than for the G-L model.}\label{fig:2}
\end{figure}

 The cluster size distribution for both models is of the form $n_s \propto s^{-\kappa}\,e^{-\mu s}$.
The scale of $s$ at which
the exponential cut-off becomes relevant, can be identified by looking at the ratio 
\begin{equation}
\frac{n_{s+1}}{n_s}=e^{-\mu}\,\frac{(s+1)^{-\kappa}}{s^{\kappa}}=
e^{-\mu}\left(1-\frac{\kappa}{s}+O\left(\frac{1}{s^2}\right)\right)\ .
\end{equation}
We assumed that $\mu \ll 1$ which is the regime in which such models exhibit power-law behavior.
The exponential cutoff becomes dominant at the scale where $a \approx (1-\frac{\kappa}{s})$, hence we may define
\begin{equation}
s_{\text{cutoff}}\equiv \frac{\kappa}{1-e^{-\mu}}\ .
\end{equation}
For the models of interest in this paper with $\mu \ll 1$, and therefore $\nu \ll 1$, we have
\begin{equation}
\text{G-L}:\;s_{\text{cutoff}}=\nu^{-1}\,,\quad \text{E-Z}:\;s_{\text{cutoff}}=\frac{5}{2}\left(\frac{2-\nu}{\nu}\right)^2\approx10\nu^{-2}\,.
\end{equation}
It is clear (see Fig.~\ref{fig:2}) that the range of cluster sizes for which one observes the power-law,
is several orders of magnitude larger for the E-Z model than for the G-L model.   
We may also verify that the special equilibrium result mentioned earlier for the continuous G-L model (see statement in italics) is also a property of the corresponding discrete model, once
the upper limits in the sums are extended to infinity. It also holds that 
\begin{equation}
\text{G-L model}:\ L_F(s)\cong s \nu L_C(s)\,,\quad\text{E-Z model}:\ L_F(s)\cong\frac{\nu}{2}L_C(s)\,.
\end{equation}

\begin{figure*}[ht]
\centering
\includegraphics[width=0.7\textwidth]{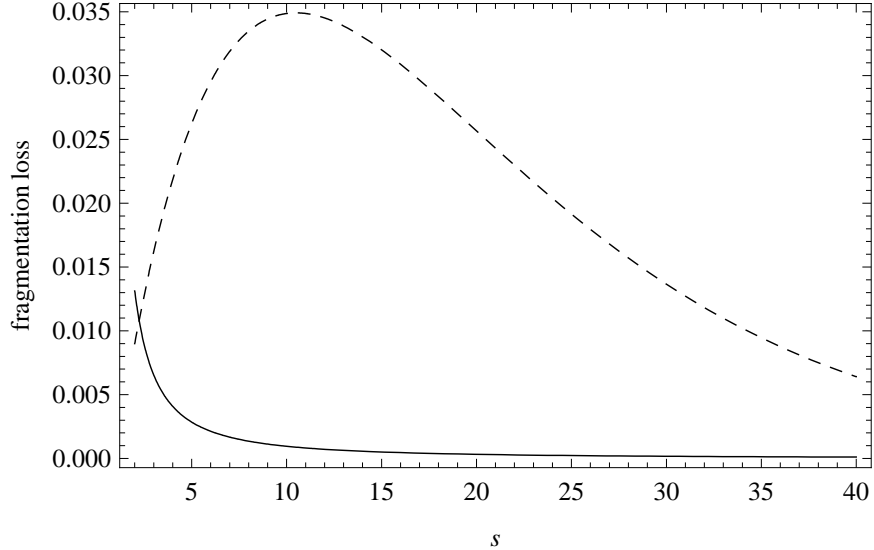}
\caption{$L_F(s)$, \textit{loss due to fragmentation} for the E-Z model (solid curve) and for the G-L model  
(dashed curve) with parameters $\nu=0.1$ and $N=1000$. The overall scale is determined up to a multiplicative constant (i.e. the scale of time). The graphs show that $L_F(s)$ for the G-L model is usually much larger 
than $L_F(s)$ for the E-Z model.
}\label{fig:3}
\end{figure*}

\noindent We see therefore that for the G-L model we can always find a value of $s$ for which $L_F(s)\approx L_C(s)$ -- in particular, it is the scale of
the cluster size over which the exponential cutoff becomes apparent. By contrast, in the E-Z model for $\nu^{-1}\gg 1$ (i.e. for the wide
range over which there is power-law behavior) we have $L_F(s) \ll L_C(s)\approx G_C(s)$.
If we again compare both models, we find that the $L_F(s)$ function is usually much larger for the G-L model 
than for the E-Z model.
Figure \ref{fig:3} illustrates this finding for a particular set of parameters.

\subsection{Reservoir model. Fragmentation into clusters of fixed size}
So far we have  looked at
the cases in which the total population size $N$ is treated as one
of the parameters defining the model. Specifically, we considered a constant population such that the constraint
$N=\sum_{s=1}^{N}sn_{s}(t)$ strictly holds at every instant in time. This highly idealized situation might not be realized in a particular real-world problem -- however we can formulate a `reservoir model' version in which the total population size is no longer a parameter defining the model,  but instead becomes a dynamical
variable whose averaged equilibrium value is determined by the model itself: $N\equiv\sum_{s=1}^{\infty}s\langle n_{s}(t)\rangle$. 
We introduce a constant supply of individuals from a system reservoir, with  $\gamma$ 
denoting the rate at which single individuals are added. 
The products of the fragmenting cluster are then moved back to the reservoir. An equivalent interpretation is that 
a cluster stays in the system but ceases to interact (i.e. it does not merge with other clusters).

Here we discuss the case which in the remainder of the dynamics resembles the terms in the E-Z model, with $\beta s$ being the rate of removing a cluster of size $s$
and $\alpha s s'$ being the coalescence rate. This particular reservoir model 
is therefore described by three parameters $\alpha,\beta,\gamma$, with only 
two parameters required for the steady-state cluster size distribution. The master equations are
\begin{equation}
-\beta s\,\ns-\alpha\,s\,\ns \sum_{s'=1}^\infty\,s' \nsp +\half \alpha\,\sum_{s'=1}^{s-1} \,s'\,\nsp \,(s-s')\nssp +
\gamma\delta_{s1}=0\,.\label{meqrm}
\end{equation}
By summation of Eq. \eqref{meqrm}, the average number of participants is obtained as 
\begin{equation}
\langle N \rangle=\frac{\sqrt{\beta^2+2\alpha\gamma}-\beta}{\alpha} \, \ .
\end{equation}
The cluster size distribution has the same form as for the E-Z model Eq. \eqref{EZ1}, if expressed in terms of $\langle N \rangle$
and $\alpha/\beta$. In this case there is no approximation made in extending the summation limit to infinity, and the solution in Eq. \eqref{EZ1} 
is exact from the mean-field theory point of view. There is no limit on the maximum size of a cluster, which in principle may 
exceed $\langle N \rangle$ when the effect of fluctuations is non-negligible.

We note that the E-Z model changes very little if we consider the case where
a cluster fragments into a set of  smaller clusters, 
each of fixed size $s_0$. For the discrete system, there is 
naturally a divisibility problem regarding fragmentation of
clusters of sizes which are not a multiple of $s_0$. Since we are interested in 
steady-state behavior, we may assume that such clusters do not fragment. Whatever 
the initial configuration is after a sufficiently long time, the system in equilibrium
will consist almost entirely of clusters that are a multiple of $s_0$ in size. 
It turns out that the cluster size distribution has the same form as the E-Z model in Eq. \eqref{EZ1}, if we re-express it in terms of $s_0$ as the basic unit,
i.e. if we substitute $s\rightarrow s / s_0$.

\section{Generalization of the E-Z model\label{sec:Generalization}}
We now open up the above
discussion to a broader class of coalescence-fragmentation models. The variety of coalescence-fragmentation-type processes
which have been employed to describe physical, biological and
social systems in the literature is enormous
\cite{fjmo,ez,gl,marker2,marker3,mhj1,mhj2,marker5,
marker6,marker7,marker8,marker9,marker10,marker13,marker14}. Here we focus on the E-Z model \cite{ez} given its potential relevance to understanding the empirical distributions observed in financial markets and insurgent behavior \cite{gabaix,neil1,clauset1,neil2,clauset2}. In particular, we will investigate the effect of variations in the rules, and perturbations, on the cluster size distribution.

\subsection{Spontaneous cluster formation}\label{prtb}
Our first generalization mimics the situation in which a small number of clusters are allowed to spontaneously form from the population, as opposed to arising from the merger of two smaller clusters. In practice this is most simply viewed as the spontaneous formation of clusters from previously single agents/clusters of unit size. (The exact mechanism is unimportant). Let $\gamma_s$ represent the rate of formation of clusters of size $s$ by the non-hierarchical method. The value of $\gamma_1$  is implicitly defined by the requirement that the size $N$ of the population remains constant, i.e., $\sum_{s=1}^{\infty} s \gamma_s = 0$, therefore
$\gamma_1<0$.
The rate equation is given by
\begin{eqnarray*}
\frac{\partial n_s}{\partial t} &=& - \beta s n_s  \nonumber\\
&&+ \half \alpha \sum_{r=1}^{s-1} rn_r\left( s-r\right) n_{s-r}  \nonumber\\
&&- \alpha sn_s   \sum_{r=1}^{\infty} rn_r + \gamma_s
\end{eqnarray*}
for $s \geqslant 2$, and
\begin{eqnarray*}
\frac{\partial n_1}{\partial t} =- \alpha n_1   \sum_{r=1}^{\infty} rn_r +\beta \sum_{r=1}^{\infty} r^2 \,n_r+ \gamma_1
\end{eqnarray*}
for $s =1$.
In the steady state this may be written as
\begin{equation} \label{eq:2.2}
sn_{s} = A{\Bigg( } \half \alpha \sum _{r=1}^{s-1}rn_{r} \left(s-r\right)n_{s-r}  \nonumber \\ + \gamma_s {\Bigg )},
\end{equation}
where $A$ is defined by
\begin{equation*}
A = \frac{1}{\beta + \alpha \sum _{r=1}^{\infty}rn_r}.  
\end{equation*}
In deriving these results, we have extended the summation of appropriate low-order terms to infinity by introducing the approximation
$\sum _{r=1}^{\infty }rn_{r} \approx N$.
The generating function $g[y]$ is now introduced:
\begin{equation} \label{eq:2.4}
g[y] \equiv \sum _{r=2}^{\infty}rn_r y^r .
\end{equation}
Taking the square of this function and using Eq.~(\ref{eq:2.2}) yields
\begin{eqnarray} \label{eq:2.5}
0&=& \left(g[y]\right)^2 -2\left(\frac{1}{A \alpha} -n_1 y\right)g[y] \nonumber \\
&& + n_1^2 y^2 + \frac{2}{\alpha} \chi[y] ,
\end{eqnarray}
where $\chi [y]\equiv \sum_{r=2}^{\infty}\gamma_r y^r$.
Using the fact that $g[1]=\sum _{r=1}^{\infty}rn_r  -n_1 $, gives
\begin{equation} \label{eq:2.6}
n_1 = \frac{1-A^2\left(\beta^2 +2\alpha \chi[1]\right)}{8A \alpha^{-1}} .
\end{equation}
Solving Eq.~(\ref{eq:2.5}) for general $y$ and expanding the resulting radical using Taylor's theorem yields
\begin{eqnarray} \label{eq:2.7}
g[y] &=& A\chi[y] + \frac{1}{A\alpha} \sum_{k=2}^{\infty} {\bigg(}\frac{(2k-3)!!}{(2k)!!} \nonumber \\
&& \times \left[2A\alpha\left(n_1 y+A \chi[y]\right)\right]^k {\bigg)} .
\end{eqnarray}
We will assume that the gamma term is small enough to be treated as a perturbation, i.e. $\frac{A \chi [y]}{n_{1} y} \ll 1$ and hence a first-order binomial expansion of the exponential term in Eq.~(\ref{eq:2.7}) may be performed. In this case
\begin{eqnarray*}
g[y] & \approx & A \chi [y] \nonumber \\
&& + \frac{1}{A\alpha} \sum_{k=2}^{\infty} {\Bigg(} \frac{(2k-3)!!}{(2k)!!} \nonumber \\
&& \times {\bigg\{}\left[ 2A\alpha n_{1} y\right]^{k} \nonumber \\
&&  + k\frac{A }{n_{1} } \left[2A\alpha n_{1} \right]^{k} \sum _{r=2}^{\infty }\gamma _{r} y^{r+k-1}  {\bigg\}} {\Bigg)} .
\end{eqnarray*}
Comparing terms with Eq.~(\ref{eq:2.4}) yields
\begin{equation*}
n_{2} =\frac{1}{2} A \gamma _{2} +\frac{1}{4}  A \alpha (n_{1} )^2
\end{equation*}
for $s=2$. For large $s$, Stirling's approximation yields
\begin{widetext}
\begin{eqnarray} \label{eq:2.14}
n_s & \approx & \left(\frac{e^2 }{2\sqrt{\pi} A \alpha} \right) 
 \left\{1-A^2 \left[\beta^2 +2\alpha \mathrm{X} \right]\right\}^s s^{-5/2}  \nonumber \\
 && + A {\bigg[}\gamma_s +\frac{e^2 }{2\sqrt{\pi} n_1 }
 \sum_{r=2}^{s-1}\left( \left\{1-A^2 \left[\beta^2 +2\alpha \mathrm{X} \right]\right\}^r r^{-1/2} \gamma_{s-r+1} \right) {\bigg]} s^{-1} ,
\end{eqnarray}
\end{widetext}
where $\mathrm{X} \equiv \chi [1]$.
Since $A$ is constant for a given population, the general form of the above equation is
\begin{equation} \label{eq:2.15}
n_s \propto \kappa^s s^{-5/2} + Z[s] s^{-1},
\end{equation}
where $\kappa\equiv 1-A^2 (\beta^2 +2\alpha \mathrm{X} )$ and $Z[s]$ is a function whose form depends on the details of the perturbation induced by the $\gamma_s$ terms.

\subsection{Step perturbation\label{sec:2-popn}}

\begin{figure*}
\includegraphics[width=0.7\textwidth]{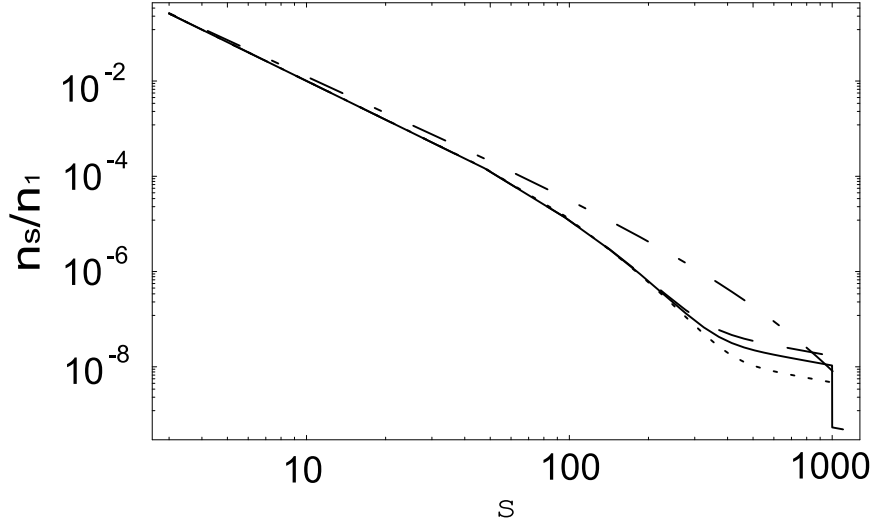}
\caption{Predicted distribution  of cluster sizes for the perturbed system described in Section \ref{sec:2-popn}, using $N=1000$, $\nu=0.1$ and $\Phi=0.01$. The dot-dashed line shows the unperturbed population, while the dashed line shows $q=10$, the dotted line shows $q=100$, and the solid line shows $q=1000$.}\label{fig:steppred}
\end{figure*}

We now analyze a highly simplified example from the class of perturbations which die off as $s$ increases. In particular, we consider a step function perturbation:
\begin{equation*}
\gamma_s =
\begin{cases}
\frac{\Phi}{q-1} , & \text{for $2 \leqslant s \leqslant q$};\\
0 , & \text{for $s > q$} ;
\end{cases}
\end{equation*}
where $q$ is an arbitrarily chosen cluster size and $\Phi>0$.
Using the original E-Z parametrization  of Eq. \ref{ezconst} and Eq.~\ref{eq:2.14}, we obtain the cluster size distribution as  
\begin{equation*}
n_1 \approx N\frac{1-\Phi }{2-\nu} ,
\end{equation*}
\begin{equation*}
n_2 \approx \frac{1}{2(2-\nu)} \left[\frac{\Phi }{q-1} +\frac{1-\nu }{(2-\nu )^2 } (1-\Phi)^2 \right] ,
\end{equation*}
\begin{eqnarray*}
n_s &\approx &  N\frac{(2-\nu )e^2 }{4\sqrt{\pi} (1-\nu)} \left[\frac{4(1-\nu )}{(2-\nu )^{2} } (1-\Phi )\right]^{s} s^{-5/2} \nonumber \\
&& + N\frac{1}{2-\nu } \frac{\Phi }{q-1} s^{-1} \nonumber \\ 
&& + \frac{e^2}{2\sqrt{\pi}} \frac{\Phi }{(1-\Phi )} \frac{1}{q-1} \nonumber \\
&& \times \left\{\sum _{r=2}^{s-1}\left[\frac{4(1-\nu )}{(2-\nu )^{2} } (1-\Phi )\right]^{r} r^{-1/2}  \right\}s^{-1} ,
\end{eqnarray*}
for $3 \leqslant s \leqslant q$, and
\begin{eqnarray*}
n_s &\approx &  N\frac{(2-\nu )e^{2} }{4\sqrt{\pi } (1-\nu )} \left[\frac{4(1-\nu )}{(2-\nu )^2 } (1-\Phi )\right]^s s^{-5/2} \nonumber \\
&& + \frac{e^2}{2\sqrt{\pi}} \frac{\Phi}{(1-\Phi)} \frac{1}{q-1} \nonumber \\
&& \times \left\{\sum_{r=2}^q\left[\frac{4(1-\nu)}{(2-\nu)^2} (1-\Phi)\right]^r r^{-1/2} \right\}s^{-1}
\end{eqnarray*}
for $s \geqslant q+1$.
Examples of the resulting $n_s$ distribution are plotted in Fig.~\ref{fig:steppred}.

\begin{figure}
  \includegraphics[width=0.7\textwidth]{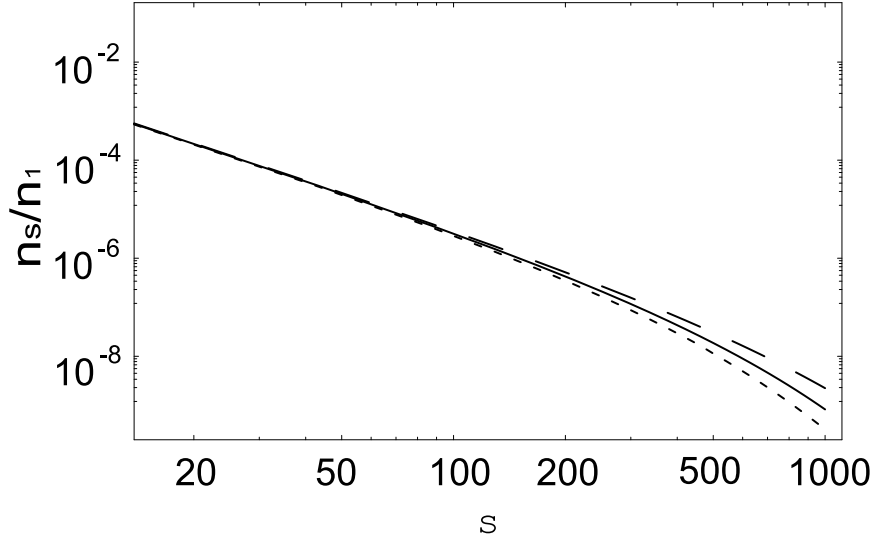}
  \caption{\label{fig:m_vs_n} Predicted cluster size distribution for the $+\Phi$ case (dotted) and the $-\Phi$ case B (dashed) compared with the unperturbed model (solid). Parameter values: $N=10000$; $\Phi=0.001$; $\nu=0.1$; $q=500$.}
\end{figure}

Interestingly the greatest effect of the perturbation is found at high $s$, whereas the perturbation's definition means that it only directly affects the clustering at low $s$. This is because the perturbation creates small clusters by non-hierarchical means, which then serve as effective nucleation sites for the formation of larger clusters. The perturbation therefore greatly accelerates the formation of large clusters whereas, by contrast, the small clusters fragment sufficiently fast that their presence is hidden on the graph at low $s$.
Figure~\ref{fig:m_vs_n} shows the predicted distribution of $n_s$ for different signs of the perturbation ($\pm\Phi$), together with the unperturbed result. 
Note that in the case of a negative sign, it is necessary that
\begin{equation} \label{eq:3.7} 
\Phi < \frac{\nu^2}{4\left(1-\nu \right)} 
\end{equation}
in order that $n_s$ remain finite as $s \to\infty$.
The analytic predictions for the perturbed populations are quantitatively reliable for a wide range of $s$ values. With primed quantities referring to the $-\Phi$ case, and using$N=10000$, $\Phi =0.001$, $\nu = 0.1$ and $q=500$, we find that the effect of the perturbation is as follows:
\begin{eqnarray*}
\frac{n_1}{n'_1} &=& 0.998 , \\ 
\frac{n_{500}}{n'_{500}} &=& 0.39 , \\ 
\frac{n_{1000}}{n'_{1000}} &=& 0.14 .
\end{eqnarray*}
As claimed earlier, this small, low-$s$ perturbation can be seen to have a very significant effect across a wide range of $s$, in particular at high $s$. We note that the interpretation of the perturbation is that statistically a cluster of size 500 or less spontaneously forms/fragments for $+/-\Phi$ cases respectively once in every 1000 timesteps, where a single timestep corresponds to any particular  fragmentation or coalescence event in the system.

\subsection{Variable population size \label{sec:variable}}
Only a small subset of real-world problems correspond to populations with a fixed size $N$, or with a fixed time-averaged size $N$. In this section of the paper, we develop an analytic treatment of a model which is analogous to the E-Z model, but which treats the case of a population whose size varies with time according to a simple law. 
As mentioned earlier, several real-world systems seem to have power-law behavior with exponent around 2.5. which is the same behavior as the unperturbed E-Z model -- for example, the distributions of size of trades in markets, and the size of attacks in conflict and terrorism\cite{neil,neil1,clauset1,neil2,clauset2,newman}. Such real-world observations could therefore conceivably be attributed to the E-Z model -- however this identification would be far more believable if the E-Z model's assumption of constant $N$ did not have to be made. It is known that as the years pass in an active war, an insurgent population will generally increase in size as previously passive people become recruited. Likewise as a market grows, previously inactive individuals tend to join the trading. Hence a model with increasing $N$ (or decreasing $N$ for mature wars or markets that are dying off) is of interest. Real-world examples of declining populations are also known \cite{saavdera},~\cite{kaivan}.

We now look at a version which can be treated analytically under the assumption that the coalescence processes are negligible. Although this makes it arguably more restricted than our previous versions, the advantage is that the equation retains linear temporal dynamics and admits a novel solution. Including all coalescence terms would make it non-linear, and intractable.

Our model considers a population containing $N[t]$ agents instantaneously divided into $M[t]$ clusters, as in the E-Z model. First we focus on the number of agents increasing in time, and introduce the following E-Z-like rules:
\begin{enumerate}
\item In a single timestep, with probability $p[t]$, $L[t]$ new agents are added to a single cluster of size $s$, the cluster being selected with probability proportional to $s$.
\item Alternatively, with probability $q[t] = 1-p[t]$, a randomly selected cluster fragments (selection of this cluster is independent of cluster size).
\end{enumerate} 
If the change in the number of agents is negative, then the model runs as follows:
\begin{enumerate}
\item In a single timestep, with probability $p[t]$, $L[t]$ agents are removed from a single cluster of size $s$, the cluster being selected with probability proportional to $s$. If the selected cluster has $s < |L[t]|$ then nothing occurs.
\item Alternatively, with probability $q[t] \equiv 1-p[t]$, a randomly selected cluster fragments, with selection of this cluster independent of cluster size.
\end{enumerate}
The rationale for adding or subtracting from a single cluster is that in many situations of interest, only a single cluster  will likely be involved in an external event which changes the population size. As with all these generalizations, more realistic rules can of course be explored -- but one runs the risk of obtaining increasingly complicated results.

\subsubsection{Increasing population size: $L[t]>0$}\label{sec:increasing}
The model proposed above leads to the rate equations
\begin{eqnarray*}
\dfrac{\partial n_s}{\partial t} = &\dfrac{p[t]}{N[t]} \left( (s-L[t]) n_{s-L[t]} - sn_s \right) & \!\!\!- \dfrac{q[t]}{M[t]} n_s \\ && \;\text{for $s>L[t]$}, \\
\dfrac{\partial n_s}{\partial t} = & -\dfrac{p[t]}{N[t]} sn_s - \dfrac{q[t]}{M[t]} n_s & \;\text{for $2 \leqslant s \leqslant L[t]$}, \\
\dfrac{\partial n_1}{\partial t} = &-\dfrac{p[t]}{N[t]} n_1 + \dfrac{q[t]}{M[t]} \sum_{r=2}^{\infty} r n_r & \;\text{for $s=1$},
\end{eqnarray*}
with resulting totals
\begin{eqnarray}
\frac{\text{d} N}{\text{d} t} &=& p[t] L[t] , \label{eq:fullMpos}\\
\frac{\text{d} M}{\text{d} t} &=& q[t] \left( \frac{N[t]}{M[t]} - 1 \right) \label{eq:fullNpos}.
\end{eqnarray}
The solution of the above equations clearly depends on the forms of $L[t]$ and $p[t]$. As a simple example, we take both to be constant: $L[t] \equiv L$ and $p[t] \equiv p$ for all $t$. In this case, it can be seen that for times $t \gg \frac{N[t=0]}{pL}$, Eq.~(\ref{eq:fullMpos}) yields the linear solution
\begin{equation*}
N[t] = pLt .
\end{equation*}
If we assume a similar asymptotically linear form for $M[t]$ at large $t$, $M[t] = \sigma t$, we can go on to deduce from Eq.~(\ref{eq:fullNpos}) that
\begin{equation*}
\sigma = \frac{q}{2} \left( \sqrt{4 \frac{p}{q} L + 1} -1 \right).
\end{equation*}
We now assume a linear form for all $n_s$: $n_s[t] = c_s t$. In this case, one obtains the solution
\begin{equation*}
c_1 = \frac{q}{\sigma} \frac{L}{L+1} \sum_{k=1}^{\infty}(1+kL)c_{1+kL} = \frac{q}{\sigma} \frac{L}{L+1} \left(p L-c_1\right)\,.
\end{equation*}
Therefore
\begin{eqnarray}
c_1 &=& \frac{pq}{\sigma} \frac{L^2}{L(1+q/\sigma)+1} , \nonumber \\ 
c_{1+kL} &=& \frac{pq}{\sigma} \frac{L^2}{L(1+q/\sigma)+1} \frac{\rho !^{(L)} (1+(k-1)L) !^{(L)}}{(\rho + kL)!^{(L)}} \label{eq:c_s},
\end{eqnarray}
for $k=1,2,3,$\dots, where
\begin{equation*}
\rho \equiv \left( \frac{q}{\sigma} +1 \right) L + 1,
\end{equation*}
and we have used the multifactorial function, defined by
\begin{equation*}
m!^{(n)} =
\begin{cases}
  1, & \text{if $0 \leqslant m < n$};\\
  m(m-n)!^{(n)} , & \text{if $m \geqslant n$}.\\
\end{cases}
\end{equation*}
Clearly $c_s = 0$ for $s \neq 1+kL$. Via a generalization of Stirling's approximation, 
\begin{equation*}
\ln (n!^{(b)}) \sim \frac{1}{b} ( n \ln n - n).
\end{equation*}
Applying this to Eq.~(\ref{eq:c_s}), we obtain our solution:
\begin{equation}\label{cce}
c_{1+kL} \approx \frac{pq}{\sigma} \frac{L^2}{L(1+q/\sigma)+1} \frac{e^{(L-1)/L} \rho^{\rho /L} (1+(k-1)L)^{k-1+1/L}}{(\rho+kL)^{k+\rho/L}}
\end{equation}
for integer $k \geqslant 1$.
If we take a snapshot of this system at any given time, the observed cluster size distribution will be given by Eq.~\eqref{cce}, modulo a multiplicative constant which grows linearly with time. The leading $k$-dependent behavior of Eq.~(\ref{eq:c_s}) is
\begin{equation}
\frac{\left(kL+(1-l)\right)^{1-L}}{\left(kL+\rho\right)^\rho}\left(kL\right)^{-(\rho+L-1)}\,.
\end{equation}

\subsubsection{Decreasing population size: $L[t] < 0$\label{sec:decreasing}}
For simplicity in the following analysis, we do not allow complete annihilation of clusters (i.e. we do not allow the removal of all of a cluster's members from the population). The rate equations for $L[t] < 0$ are as follows:
\begin{eqnarray*}
\dfrac{\partial n_s}{\partial t} &=& \dfrac{p[t]}{N[t]} \left( (s+|L[t]|) n_{s+|L[t]|} - sn_s \right) - \dfrac{q[t]}{M[t]} n_s \\&& \hspace{4cm}\text{for $s>|L[t]|$}, \\
\dfrac{\partial n_s}{\partial t} &=& \dfrac{p[t]}{N[t]} (s+|L[t]|)n_{s+|L[t]|} - \dfrac{q[t]}{M[t]} n_s \\&&\hspace{4cm}\text{for $2 \leqslant s \leqslant |L[t]|$},
\end{eqnarray*}
and
\begin{equation*}
\dfrac{\partial n_1}{\partial t} = \dfrac{p[t]}{N[t]} (1+|L[t]|)n_{1+|L[t]|} + \dfrac{q[t]}{M[t]} \sum_{r=2}^{\infty} r n_r
\end{equation*}
for $s=1$, with resulting totals
\begin{eqnarray}
\frac{\text{d}N}{\text{d} t} &=& -\frac{p[t] |L[t]|}{N[t]} \sum_{r=1+|L|}^{\infty} \! rn_r , \label{eq:fullMneg}\\
\frac{\text{d} M}{\text{d} t} &=& q[t] \left( \frac{N[t]}{M[t]} - 1 \right) \label{eq:fullNneg}.
\end{eqnarray}
As above, we can obtain a solution by assuming that $p$ and $L$ are both constant, and then introduce a linear trial solution of the form
\begin{eqnarray*}
N[t] &=& N_0 -\gamma t , \nonumber \\
M[t] &=& M_0 + \sigma t , \nonumber \\
n_s[t] &=& C_s - c_s t .
\end{eqnarray*}
This approximation can only hold as long as the changes in each $n_s$ are small compared to the size of the respective $C_s$. In this case (i.e., for $t$ not too large) we obtain
\begin{eqnarray*}
\sigma \approx q \left( \frac{N_0}{M_0} -1 \right) , \nonumber \\
\gamma \approx \frac{p |L|}{N_0} \sum_{r=1+|L|}^{\infty} \! rC_r ,
\end{eqnarray*}
and for the $c_s$ we obtain:
\begin{eqnarray*}
  c_1 &\approx& -\dfrac{p}{N_0} (1+|L|)C_{1+|L|} - \dfrac{q}{M_0} \left( N_0 - C_1 \right) \\ && \hspace{5.3cm}\text{for $s=1$}, \\
  c_s &\approx& \dfrac{q}{M_0} C_s -\dfrac{p}{N_0} \left(s+|L| \right) C_{s+|L|} \hspace{1cm}\text{for $2 \leqslant s \leqslant |L|$},\\
  c_s &\approx& \left( \dfrac{q}{M_0} + \dfrac{p}{N_0} s \right) C_s - \dfrac{p}{N_0} \left( s+|L| \right) C_{s+|L|} \\&& \hspace{5.3cm}\text{for $s > |L|$.}
\end{eqnarray*}
With a suitable choice of initial conditions and a large population, one can therefore infer the small-$t$ behavior of the system.

\subsubsection{Decreasing population: proof of concept\label{sec:example}}
\begin{figure}
\includegraphics[width=0.7\textwidth]{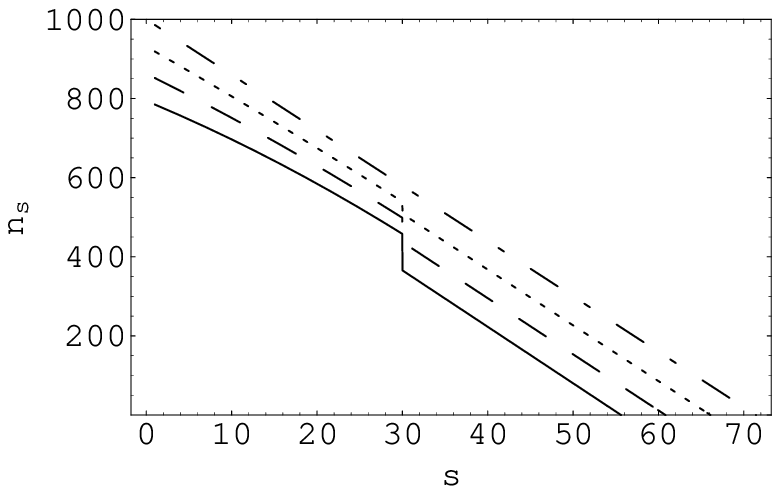}
\caption{\label{fig:variable} Predictions of the model of Section~\ref{sec:example}, using parameter values $C_1=1000$, $\phi=14$, $p=0.3$ and $L=-30$. This yields population size parameters of $N_0=850173$ and $M_0 =35214$. Line styles reflect different values of the parameter $t$: dot-dashed $(t=0)$, dotted $(t=5000)$, dashed $(t=10000)$ and solid $(t=15000)$. Beyond $t=15000$ it can be seen that the approximations made in the derivation of Section~\ref{sec:decreasing} become inaccurate.}
\end{figure}

As a simple example, we take $L<0$ and a starting population of the form
\begin{equation*}
n_s[t=0] =
\begin{cases}
  C_1-\phi s, & \text{if $s< \frac{C_1}{\phi}$};\\
  0 , & \text{if $s \geqslant \frac{C_1}{\phi}$}.\\
\end{cases}
\end{equation*}
In this case our equations from Section \ref{sec:decreasing} yield
\begin{eqnarray*}
N_0 &=& \frac{1}{6} C_1 \left[ \left( \frac{C_1}{\phi} \right)^2 -1 \right] , \nonumber \\
M_0 &=& \frac{1}{2} C_1 \left( \frac{C_1}{\phi} -1 \right) ,\nonumber \\
\gamma &\approx& p \left|L \right| \left\{ 1- \frac{|L|\left(1+|L|\right) \left( 3C_1 -\phi -2|L|\phi \right)}{C_1 \left[ \left(\frac{C_1}{\phi} \right)^2 -1 \right]} \right\} , \nonumber \\
\sigma &\approx& \frac{(1-p)}{3} \left( \frac{C_1}{\phi} -2 \right) .
\end{eqnarray*}
This leads to an expression for $n_1$ of the form
\begin{eqnarray*}
n_1[t] &\approx& C_1 + \left[ \frac{p}{N_0} \left(1+\left|L\right|\right)\left(C_1 -\phi\left(1+\left|L\right|\right)\right) \right. \nonumber \\
&& \left. + \frac{q}{M_0} \left( N_0 -C_1 \right) \right] t ,
\end{eqnarray*}
with corresponding $n_s$ of the form
\begin{eqnarray*}
n_s[t] &\approx& C_1 -\phi s -\left[\left(\frac{q}{M_0} C_1 - \frac{|L|p}{N_0} C_{|L|} \right) \right.\nonumber \\
&& \left. - \left( \frac{p}{N_0} C_{2|L|} + \frac{q}{M_0} \phi \right) s + \frac{p}{N_0} \phi s^2 \right] t
\end{eqnarray*}
for $2 \leqslant s \leqslant \left| L \right|$, and
\begin{eqnarray*}
n_s[t] &\approx& C_1 -\phi s -\left[\left(\frac{q}{M_0} C_1 - \frac{|L|p}{N_0} C_{|L|} \right) \right. \nonumber\\
&& \left. + \left( \frac{2\left|L\right|p}{N_0} - \frac{q}{M_0} \right) \phi s \right] t
\end{eqnarray*}
for $s > \left| L \right|$. Figure~\ref{fig:variable} shows a plot of this model using illustrative parameter values.

\subsection{Heterogeneity of members\label{modfs}}

In many real-world systems -- in particular, biological or social systems -- the population is heterogeneous. In addition to
the basic question of whether approximating a heterogeneous system by a homogeneous model is justifiable,  there is the deeper issue of how to formally introduce heterogeneity into coalescence-fragmentation systems. As we have seen in this paper, small changes in
coalescence-fragmentation rules can sometimes yield dramatic changes in the cluster size distribution, and vice versa. In other words, the `devil may be in the detail' in terms of the emergent phenomena that can be expected from a given set of microscopic rules. Our limited goal here is to explore some encouraging developments in this area, highlighting the circumstances in which the heterogeneity of the population allows an accurate description in terms of an effective homogenous model. 

Reference \cite{marker10} introduces a `character' to each object by means of an   
$m$-dimensional normalized vector which is  formed from $m$-bit binary strings. The scalar product of any two such characters then becomes the argument of a
function which controls the coalescence and fragmentation processes. The general case requires numerical simulation. Interestingly, however, this model produces a power-law over part of its range with a $2.5$ slope which is identical to the homogenous E-Z model. Instead of the power-law exponent, it is the form of the exponential cut-off which turns out to depend on the heterogeneity of the population. We recently explored another type of heterogeneous E-Z-like model, showing that it can bridge the gap between the power-law slope of magnitude 2.5 for clusters in the E-Z model (and hence 1.5 for price returns) and the empirical value of financial market price returns which is typically closer to 4 \cite{blazej}.
A simple version of the vector model is provided via a fascinating recent variation proposed by Hui\cite{pm} in which the heterogeneity is represented by a character parameter $\rho_k\in [0,1]$ which is
assigned to each object in the entire population, where objects are numbered by
$k=1\ldots N$. The  probability that an agent $i$ and another agent $j$ form a link (and therefore for the inequivalent clusters to which these members belong to merge) depends on the value $ | \rho_i-\rho_j |$. In principle it may be a general symmetric function $p(\rho_i-\rho_j )$.
The fragmentation of a cluster may also depend on the characters of the members that form the particular cluster. One way of introducing this is by a mechanism in which fragmentation of the whole cluster 
is triggered by breaking any single link that belongs to it~\cite{pm}. Since a weaker link is easier to break, it is assumed that 
the probability that the link breaks is proportional to $p(\rho_i-\rho_j )$ which may be interpreted as a measure 
of the strength of the link formed between members $i$ and $j$. If $p(\rho_i-\rho_j )$ is a function which is sharply peaked at $0$, 
we will have a situation where the newly formed clusters consist only of members of very similar character, and the whole system
may be considered as a mixture of several homogeneous population subsystems which do not interact which each other.
Each of these subsystems is described by the cluster size distribution of the form in Eq. \eqref{EZ1} with 
constants determined by the distribution of characters across the population.
The cluster size distribution for the whole system (regardless of the character) is then a sum of the distributions for 
the subsystems -- therefore we still observe a scale-free behavior with variation in the form of the cut-off (i.e. diversity 
in the heterogeneity of the population induces diversity in the constants describing the subsystems, and hence lengthens the tail of the cluster size distribution).
In the opposite limiting case, the function $p(\rho_i-\rho_j )$ does not vary sharply over its argument, 
e.g. $p(\rho_i-\rho_j )\propto 1-|\rho_i-\rho_j|$, thereby yielding homogeneous mixing. The
the distribution of characters across different clusters is uniform and the system can therefore be described as an effectively homogeneous one
 by Eqs. \eqref{ezmeq}
and~\eqref{EZ1}. The presence of the heterogeneity changes only the value of $\alpha/\beta$ in Eq. \eqref{ezmeq}.

\section{ Conclusions and Implications}
We have examined various coalescence-fragmentation systems, with the goal of elucidating how subtle changes in their underlying rules can affect the resulting distribution of cluster sizes. In the process, we have managed to connect the rules of coalescence and fragmentation with terms in the corresponding rate equations, and have identified the specific ways in which they affect the resulting distribution of cluster sizes. The connections are not always direct, but we have offered various insights which help establish a more direct link. In each case studied, the system senses the fragmentation function in two ways: the appearance of new clusters coming from the fragments of the fragmented cluster (represented by $G_F(s)$), and
the disappearance of clusters that fragment (represented by $L_F(s)$).

As a result of our analysis, we can better understand what factors
dictate when a power-law is likely to emerge, and what tends to
control its exponent. We conclude that: (1) it is the substantial
contribution of $L_F(s)$ in the equilibrium condition (Eq.
\ref{eqcon}) which may prevent the size distribution from showing
a power-law behavior. (2) The presence or absence of $G_F(s)$
(i.e. the appearance of fragmentation products of new clusters)
influences strongly the value of the power-law exponent itself, in
cases where the power-law emerges. In the case where the parameter
controlling the fragmentation is small but finite, it is hard to
identify a common limiting case for the various systems studied --
however, the form of the fragmentation function does influence the
cluster size distribution regardless of the value of this
parameter. Note that if the fragmentation rate tends to zero, the
system cannot be clearly described using mean-field theory, since it
performs quasi-oscillatory behavior associated with the build-up of
one supercluster containing essentially the whole population, and this
supercluster's eventual break-up. Whatever the mode of
fragmentation,  the exponent of the power-law may be controlled by
altering the power of the cluster size $s$ which is involved in
the fragmentation and coalescence function. Specifying it
realistically requires some detailed understanding of the system at the microscopic
level. The most common mechanism of coalescence is created by
building random links between the population members, yielding a
coalescence function of the form $\sim s s'$.

If we adopt a point of view in which the system is considered as
an evolving network, the clusters represent disconnected
components. Depending on the particular rules, the fragmentation process now corresponds to breaking links. If the
disconnected component in a network breaks predominantly into
single members, it might be still interpreted in terms of the
fragmentation being triggered by a single member, provided we
allow some kind of link-breaking virus to spread rapidly
throughout the entire disconnected component. Somewhat
counter-intuitively, we have also seen that the behavior of the
heterogeneous system does not substantially differ from the
behavior of the homogeneous one. This results from two effects:
the homogeneous mixing effect, and the coexistence of several
non-interacting populations whose distinct `characters' lie hidden
in the cluster size distribution.

Although we have mentioned various possible applications, we
finish by noting a new one. Many of the neurodegenerative
disorders associated with aging, for example Alzheimer's disease,
are thought to be associated with the large-scale self-assembly of
nanoscale protein aggregates in the brain \cite{marker14}.
Protein-aggregation has of course attracted much attention over
the years in both the chemistry and physics literature -- however,
the problem of protein aggregates in neurodegenerative diseases is
known to be much harder than traditional polymer problems, because
of the complexity of the individual proteins themselves
\cite{marker14}. Given the wide range of possible heterogeneities
{\em in vivo} within a cell, there is typically insufficient
knowledge to specify either (i) a specific diffusion model and its
geometry and boundary conditions, as a result of geometrical
restrictions and crowding effects\cite{schnell1}, or (ii) a
specific reaction model for the binding rates, given the wide
variety of conformational states in which molecules may meet. It
therefore makes sense to assign some probabilities to the
aggregation process -- and in particular, coalescence and
fragmentation probabilities to describe the joining of an $n$-mer
with an $n'$-mer to give an $n''$-mer, where $\{n, n', n''\}\equiv
1, 2, 3, . . .$, and its possible breakup. The precise details of
the coalescence and fragmentation rules now
takes on a critical importance, since subtle changes in these rules can alter the resulting size distribution of the
$n$-mer population. The practical question of how fatal a given realization of the disease will be in a particular patient, becomes intertwined with the question of whether the distribution of cluster sizes is a regular one in terms of its fluctuations -- e.g. a Gaussian or Poisson distribution which both have a finite variance --  or it is a power-law which may then have a formally infinite variance. Although in
practice a cut-off always exists, a power-law with an exponent
$\alpha <2$ has (in principle) an infinite mean and infinite
standard deviation; a power-law with $2< \alpha <3$ has (in
principle) a finite mean but an infinite standard deviation; and a
power-law with $\alpha >3$ has a finite mean and finite standard
deviation. The implication is that a coalescence-fragmentation
process producing a power-law with $\alpha<3$ as in E-Z-type
models where $\alpha\sim 2.5$, has a significant probability of
forming very large $n$-mers because of its (in principle) infinite
standard deviation. Suppose for the moment that an $n$-mer of size
$n\geq n_0$ can produce a neurodegenerative disorder, then the
fraction of such dangerous $n$-mers in a soup of self-assembling
polymer aggregates, will be non-negligible if $\alpha <3$.  In the
highly crowded, heterogeneous $n$-mer population expected in the
human body, the resulting value of any approximate power-law slope
$\alpha$ could therefore be a crucial parameter to estimate. The
possibility of engineering this $\alpha$ value such that large
aggregates are unlikely, through subtle changes in the coalescence
and fragmentation processes, then takes on a very real possibility. It also adds direct medical relevance which justifies further work on this topic in the future.

\end{document}